\documentclass[12pt]{article}

\usepackage{lscape}
\usepackage{multirow}
\usepackage{amssymb,amsmath}
\usepackage{amstext}
\usepackage{amscd}
\usepackage{youngtab}
\usepackage{graphics}     
\usepackage{epsfig}
\usepackage{shadow}

\def\G{{\hspace{.3mm}\bf g\hspace{.3mm}}}
\def\GRAD{{\hspace{.3mm}\bf grad\hspace{.3mm}}}

\def\TR{{\hspace{.3mm}\bf tr\hspace{.3mm}}}

\def\gr{{\bf grad}}
\def\dv{{\bf div}}
\def\tr{{\bf tr}}
\def\gi{{\bf g}}

\def\be{\begin{equation}}
\def\ee{\end{equation}}
\def\beq{\begin{equation}}
\def\eeq{\end{equation}}
\def\bea{\begin{eqnarray}}
\def\eea{\end{eqnarray}} 

\def\eqn#1{(\ref{#1})}

\def\nn{\nonumber}

\def\coker{{\rm coker}}

\def\sideremark#1{\ifvmode\leavevmode\fi\vadjust{\vbox to0pt{\vss
 \hbox to 0pt{\hskip\hsize\hskip1em
 \vbox{\hsize3cm\tiny\raggedright\pretolerance10000
  \noindent #1\hfill}\hss}\vbox to8pt{\vfil}\vss}}}

\begin{document}
\thispagestyle{empty}

\vspace{.8cm}
\setcounter{footnote}{0}
\begin{center}
\vspace{-25mm}
{\Large

 {\bf Generalized Einstein Operator\\ Generating Functions}\\[5mm]

 {\sc \small
     D.~Cherney$^{\mathfrak C}$, 
     E.~Latini$^{\mathfrak L}$, 
     and A.~Waldron$^{\mathfrak W}$\\[4mm]

            {\em\small${}^{\mathfrak C,\mathfrak W}\!$
            Department of Mathematics\\ 
            University of California,
            Davis CA 95616, USA\\
            {\tt cherney,wally@math.ucdavis.edu}\\[2mm]
           ${}^\mathfrak{L}$ 
Department of Mathematics\\ 
            University of California,
            Davis CA 95616, USA\\
and\\ INFN, Laboratori Nazionali di Frascati, CP 13, I-00044 Frascati, Italy \\
{\tt emanuele@math.ucdavis.edu, latini@lnf.infn.it}\\[2mm]

            }}

 }

\bigskip

{\sc Abstract}\\[-4mm]
\end{center}

{\small
\begin{quote}
We present gauge invariant, self adjoint Einstein operators for mixed symmetry higher spin theories.
The result applies to multi-forms, multi-symmetric forms and mixed antisymmetric and symmetric multi-forms,
generalizing previous results for combinations of these cases.
It also yields explicit action principles for these theories in terms of their minimal covariant field
content. For known cases, these actions imply the mixed symmetry equations of motion of Labastida.
The result is based on a  calculus for handling normal ordered operator expressions built from  quantum generators
of the underlying constraint algebras.
\end{quote}
}

\noindent The dynamics of higher spin fields 
is described by the equation of motion\footnote{We refer the reader to the series of higher spin review 
articles \cite{Bekaert:2005vh}.}
\bea \label{labastida}
\Big(\underbrace{ \Delta -Q^iQ^*_i+\frac{1}{2}Q^iQ^j{\bf tr}_{ji}}_{\textstyle{\bf G}}
\Big){\bf \Psi}=0={\bf tr}_{i(j}{\bf tr}_{km]}{\bf \Psi}\, ,
\eea
which enjoys the gauge invariance
\bea \label{gauge}
\delta{\bf \Psi}=Q^k{\bf \alpha}_k\, ,~~
\eea
for gauge parameters ${\bf \alpha}_k$ subject to the trace condition
$${\bf tr}_{(ij}{\bf \alpha}_{k]}=0\, .$$
We refer to ${\bf G}$ as the Labastida operator.
It acts on  
mixed multi-forms:
$$
{\bf \Psi}\!=\!{\small {\bf \Psi}_{
\mu_1^1\ldots\mu_{s_1}^1\cdots\mu_1^q\ldots\mu_{s_q}^q\nu_1^1
\ldots
\nu_{t_1}^1\cdots\nu_1^r\ldots\nu_{t_r}^r}
d^1x^{\mu_{1}^1}\!\wedge\!\ldots\!\wedge\! d^1x^{\mu_{s_1}^1}
\!\otimes\! \cdots\! \otimes\!
d^qx^{\mu_{1}^q}\!\wedge\!\ldots\!\wedge\! d^qx^{\mu_{s_q}^q}}
$$
$$\qquad\qquad\qquad\qquad\qquad\qquad\quad\ \  
\otimes\:
d^1x^{\nu_{1}^1}\odot\ldots\odot d^1x^{\nu_{t_1}^1}
\otimes \cdots \otimes
d^rx^{\nu_{1}^r}\odot\ldots\odot d^rx^{\nu_{t_r}^r}\, ,
$$
%
or, in a Young diagrammatic notation
$$
\Yvcentermath1
{\bf \Psi}=\yng(1,1,1)\otimes...\otimes\yng(1,1,1,1,1)\ \ \otimes\ \begin{tabular}{c}\yng(4)\\ $\otimes$\\ \vdots\\ $\otimes$ \\[1mm] \ \!\! \yng(3) ~\end{tabular}~.
$$ 
{\it I.e.}, an arbitrary representation of $\frak{gl}(d)$
 constructed with $q$~columns and~$r$ rows. One can think of such tensors as monomials, or better--functions, in the super anti-commuting variables
$\eta_i^\mu\equiv d_ix^\mu$ where the superindex 
$$i=(\underbrace{1,...,q}_{\rm Fermi}|\underbrace{1,...,r}_{\rm Bose})~.$$

The (super)charges $(Q^*_i,Q^j)$ act on blocks of ${\bf \Psi}$ as exterior derivatives or symmetrized gradients and their duals
\bea
Q^i&=&(
{\bf d}^1,..,{\bf d}^q
|{\bf grad}^{1},..,{\bf grad}^ {r})\nonumber \, ,\\[3mm]
Q^*_i&=&(
{\bf d}^{*}_{1},..,{\bf d}^{*}_{q}
|{\bf div}_{1},..,{\bf div}_{r})
\nonumber\, ,
\eea
while $\Delta $ is the Laplace operator. The BRST quantization of these operators was first studied in~\cite{Henneaux:1987cp}.
The algebra of these (super)charges,
in an $d$-dimensional flat spacetime background (for studies of more general backgrounds see \cite{Hallowell:2005np,Bastianelli:2007pv,Marcus}),
$$
[Q^*_i,Q^j\}=\delta^j_i\Delta \, ,
$$
may be extended by the $R$-symmetry generators $\TR_{ij}$
$$
[\TR_{ij},Q^k\}=2\delta^k_{(j}Q_{i]} \, .
$$
The operator $\tr_{ij}$ traces across the $i^{\rm th}$ and $j^{\rm th}$ blocks of indices while Its dual $\G^{ij}$ uses the metric tensor to add an index to the $i^{\rm th}$ and $j^{\rm th}$ blocks with supersymmetry between the new indices. 
These operators generate the full $R$-symmetry algebras $\frak{osp}(q,q|2r)$ 
 which we use to label the first class algebras $\{Q^i,\Delta,\TR_{ij},Q^*_i\}$.
 
In~\cite{Cherney:2009mf} it was shown how BRST detour quantization of these algebras  produces the Labastida operators as long operators of detour complexes. The $\frak{sp}(2r)$ version of that computation was given in~\cite{Alkalaev:2008gi}.
Originally, the equation of motion~\eqn{labastida} was first given in~\cite{Labastida:1986gy} for the case $r=0$, while its multi-form analog at $q=0$ was studied in~\cite{de Medeiros:2002ge}.
The case of mixed multi-forms was first introduced in~\cite{Henneaux:2008ee,Bekaert:2002dt} and studied in detail using BRST detour quantization techniques in~\cite{Cherney:2009mf,Bastianelli:2009eh}. The above equation of motion generalizes the pure multi-form and multi-symmetric form  models to the mixed case at arbitrary~$(q,r)$. An unfolded formulation and action principle for mixed higher spins was given in~\cite{Skvortsov:2008vs}
(see also~\cite{Zinoviev:2002ye} for first order approaches to mixed symmetry fields). The main aim of this Letter
is to write action principles for all these theories in terms of a minimal covariant field content. 

\section*{Einstein Operators}

We search for action principles of form
\be\label{action}
S=({\bf\Psi},\mathcal{G}{\bf\Psi})\equiv\int \!\sqrt{-g} \  {\bf\Psi}^*\mathcal{G}{\bf\Psi}\, ,
\ee
which imply the equation of motion~\eqn{labastida}.
Here ${\bf \Psi}^*$ is defined by replacing oscillator variables by derivatives $(\eta_i^\mu)^*= \frac{\partial}{\partial \eta_i^\mu}$,
and the inner product is computed by allowing these  to act to the right on the corresponding oscillators in ${\bf \Psi}$.
We call the operator ${\mathcal G}$ appearing in the action the Einstein operator. It is required to be self-adjoint, in the sense $({\bf \Phi},\mathcal{G}{\bf \Psi})=(\mathcal{G}{\bf \Phi},{\bf \Psi})$.

Another subtlety to bear in mind is that the fields in the Labastida equation of motion obey the double trace constraint 
${\bf tr}_{i(j}{\bf tr}_{km] }{\bf \Psi}=0$. Hence, any terms in $\mathcal{G}$ of the form  ${\bf g}_{i(j}{\bf g}_{km] }{\bf X}$
will vanish in the inner product. In more mathematical terms, the Einstein operator $\mathcal{G}$ is a mapping
$$\ker ({\bf tr}_{i(j}{\bf tr}_{km]})\longrightarrow \coker\, ({\bf g}^{i(j}{\bf g}^{km]})\, .$$ 

As a warm up we analyze the case of ${\frak{sp}}(2)$ ({\it i.e.}, $q=1$, $r=1$), higher spin, totally symmetric~tensors $${\bf \Psi}={\bf \Psi}_{\mu_1..\mu_t}dx^{\mu_1}\odot \ldots\odot dx^{\mu_{t}}~.$$ The Labastida field equation ${\bf G \Psi}=0$ 
can be bought to a form that comes from the variation of an action by adding a term obtained by tracing the original equation
\be \label{example}
\Big(1-\frac 1 4\gi\, \tr \Big){\bf G \Psi}\equiv\mathcal{G}{\bf \Psi}=0\, .
\ee
In this case $\mathcal{G}$ is the  desired, manifestly gauge invariant,  and self adjoint Einstein operator\be\label{ge}
\mathcal{G}=\Delta -\gr\, \dv+\frac 1 2\Big(\gr^2\, \tr+\gi\, \dv^2 \Big)-\frac 1 4\gi\Big(2\Delta +\gr\, \dv\Big)\tr\, .
\ee
When $t=2$ we may write ${\bf \Psi}=h_{\mu\nu}dx^\mu \odot dx^\nu$ ({\it{i.e.}}, metric fluctuations) and find the linearized Einstein tensor
$$
\mathcal{G}{\bf \Psi}=\Big(\Delta h_{\mu\nu}-
2\partial^\rho \partial_{ \mu}h_{\nu \rho}+
\partial_\mu\partial_\nu h+
\eta_{\mu\nu}\partial^\rho\partial^\sigma~h_{\rho\sigma}-\eta_{\mu\nu}\Delta h   \Big)\,  dx^\mu \odot dx^\nu~.
$$
This is our motivation for calling  $\mathcal{G}$ an ``Einstein operator''\footnote{One might also be tempted by the term ``Maxwell operator'', since when $t=1$ we have $\Psi=A_\mu dx^\mu$ and equation~(\ref{ge}) reproduces  Maxwell's equations. Also, in the $\mathfrak{so}(1,1)$ case, our Einstein operator
 again describes  Maxwell's  equations and their higher form analogs  ${\bf d^*d\Psi}=0$ (in fact in any curved background, see~\cite{Bastianelli:2009eh}). }.

A similar result holds in general but, as the symmetry type of the tensors involved becomes more complicated, higher traces of the Labastida equation of motion must be used to produce an Einstein operator. For the symmetric multi-form case, these trace terms were first computed in~\cite{Campoleoni:2008jq}
(where also the idea for the antisymmetric multi-form case was outlined\footnote{Those authors have also extended those results to generalized Dirac operators describing fermionic higher spins~\cite{Campoleoni:2009gs}.}.) Earlier results for the symmetric bi-form case were given in~\cite{Burdik:2000kj}. We have found a simple formula for these terms in the most general case of mixed multi-forms based on the BRST detour methodology of~\cite{Cherney:2009vg,Bastianelli:2009eh}.
This method has the advantage that it arranges all elements of the BRST cohomology either as field equations, gauge (and gauge for gauge) invariances
or Bianchi (and Bianchi for Bianchi) identities. Our result closely mimics the one of Sagnotti {\it et al}~\cite{Campoleoni:2008jq}
 and reads
\be\label{genfct}
\mathcal{G}=\textrm{{\large :}}~\frac{I_1(2\sqrt{\omega})}{\sqrt{\omega}}~\textrm{{\large :}}~{\bf G}\, ,\ee
with
$$
\omega=-\frac{\gi^{ji}{\bf tr}_{ji}}{2}~.
$$
Here the symbols $\textrm{{\large :}}~\bullet~\textrm{{\large :}}$ indicate normal ordering $(Q^i,{\bf g}^{ij},\Delta ,{\bf tr}_{ij},Q^*_i)$ and the modified
Bessel 
function of the first kind divided by the square root of $\omega$ has the analytic expansion
\be\label{Bessel}\frac{I_1(2\sqrt{\omega})}{\sqrt{\omega}}=\sum_n\frac{\omega^n}{n!(n+1)!}\equiv B(\omega)\, .\ee
Explicitly, the generating function for the Einstein operator appearing in~(\ref{action}) reads\footnote{ 
In order to compare with the result proposed in \cite{Cherney:2009vg} note that $B+\omega B'=I_0(2\sqrt{\omega})$ Bessel function of the first kind.}
\be
\shabox{
$
\mathcal{G}=\textrm{{\large :}}\, (B+\omega B')~\Delta -Q^i B~ Q^*_i
-Q^k \gi^{mj}B'\, {\bf tr}_{mk}Q^*_j
\nonumber \\[3mm]~~~~~~~~~~~~~
+\frac 1 2\gi^{ij}B~Q^*_jQ^*_i+\frac 1 2Q^iQ^jB~{\bf tr}_{ji}
\, \textrm{{\large :}}
$}
\ee
Written in the form~\eqn{genfct}, the gauge invariance~\eqn{gauge} is manifest (since identically ${\bf G} Q^k\alpha_k=0$)
while the above form makes self-adjointness of~$\mathcal{G}$ and, consequently, the Bianchi identity 
$Q^*_k \mathcal{G} {\bf \Psi}=0$ (modulo $\G^{(ij} {\bf X}^{k]}$) manifest.   
However, it is also possible to prove directly the gauge invariance of the action~\eqn{action} because there is a {\it calculus} of the above
normal ordered operator expressions: 

\begin{center}
\shabox{
$
\textrm{{\large :}}\, f(\omega)~\textrm{{\large :}}~Q^k=\textrm{{\large :}}~Q^k f(\omega)+f'(\omega)~\gi^{ki}Q^*_i\, \textrm{{\large :}}
$
}
\end{center}
for any function $f$. This formula converts an apparently difficult algebraic computation into simple differentiations!

Finally, we note that arbitrary symmetry higher spin gauge theory presented here can also be viewed as a detour complex\footnote{We point out that in~\cite{Henneaux:2008ee} and \cite{Bekaert:2002dt} a de Rham complex for multi-form higher spin curvatures have been constructed. Our detour complex  generalizes that result to mixed  multi-forms  and is formulated in terms of {\it gauge potentials}.}
 \be\nn
\shabox{$\hspace{-.4cm}
\begin{array}{c}
\stackrel {}{\cdots \longrightarrow} 
{\ker~{\bf tr_{(ij}}}
\stackrel{\bf  Q^k} {\longrightarrow}
{\ker~{\bf tr}_{i(j}{\bf tr}_{km]}}
~~~~{ \coker~{\bf g}^{i(j}{\bf g}^{km]}}
\stackrel{\bf  Q^*_k}{\longrightarrow} 
{\coker~{\bf g}^{( ij} }
\stackrel{ }{\longrightarrow\cdots }
\\
\ \Big|\hspace{-.9mm}\raisebox{-2.5mm}{\underline{ \quad\ ~~ \raisebox{.5mm}{$ {\bf {\cal G}}$ }\quad~~   }} \hspace{-1.3mm}{\Big\uparrow}
\end{array}\hspace{-.4cm}
$}
\ee
whose physical interpretation  is the following: $\ker{\bf tr}_{i(j}{\bf tr}_{km]}$ and $\ker{\bf tr}_{(ij}\equiv \{\alpha_k : {\bf tr}_{(ij}\alpha_{k]}=0\}$ are the spaces of gauge potentials and gauge parameters
and comprise the ``incoming complex'' which encodes gauge and gauge for gauge symmetries (see~\cite{Cherney:2009mf} and~\cite{Bastianelli:2007pv} for a complete description).
The duals of these spaces, $\coker\, {\bf g}^{i(j}{\bf g}^{km]}$ and $\coker\, {\bf g}^{(ij}$, correspond to equations of motion and Noether/Bianchi identities
and comprise  the ``outgoing'' complex. The Einstein operator  is ``long operator''  joining the two.

In the following sections we discuss some simple examples. 

\subsection*{$\mathfrak{sp}(4)$: Symmetric Multi-Forms}
The $q=0,\,r=2$ case describes multi-symmetric-form gauge potentials
\vspace{2mm}
\be\label{yt1}
\Yvcentermath1
{\bf \Psi}=
{\bf \Psi}_{\mu_1^1\ldots\mu_{t_1}^1\mu_1^2\ldots\mu_{t_2}^2}
d^1x^{\mu_{1}^1}\odot\ldots\odot d^1x^{\mu_{t_1}^1}
\otimes 
d^2x^{\mu_{1}^2}\odot\ldots\odot  d^2x^{\mu_{t_2}^2}
 =\!\begin{tabular}{c} \yng(3)\\ $\otimes$ \\[1mm] \yng(4) \end{tabular}\, .
\ee
The  Bessel series~\eqn{Bessel}, when acting on $\ker{\bf tr}_{i ( j}{\bf tr}_{km )}$, terminates at second order (by virtue of the paucity of index values)  so that the generalized Einstein equation of motion is\footnote{Repairing a factor 4 typographical error in the last term relative to the result of~\cite{Bastianelli:2009eh}.}
\bea
\mathcal{G}{\bf \Psi}&=&\Big(
\Delta -{\bf grad}^i\dv_i
+\frac 1 2 \gr^i\gr^j\tr_{ji}+\frac 1 2 \gi^{ij}\dv_j\dv_i-\frac 12 \gi^{ij} ~\Delta~\tr_{ji} \nonumber\\
&&+\frac 1 4  \gr^i\gi^{km}\tr_{mk}~\dv_i-\frac 1 2{\bf grad}^k\gi^{mj}\tr_{mk}{\bf div}_j
\nonumber\\
&&
-\frac{1}{48}\gr^i\gi^{km}\gi^{pq} \tr_{qp}\tr_{mk}~\dv_i
+\frac{1}{12}\gr^k\gi^{mj}\gi^{pq}\tr_{qp}\tr_{mk}~\dv_j  \nonumber\\
&&-\frac 1 6 \gi^{i[j}\gi^{k]m}\tr_{mk}~\dv_j\dv_i-\frac 1 6 \gr^i\gr^{j}~\gi^{km}\tr_{m[k}\tr_{j]i}\nonumber\\&&+\frac{1}{16}\gi^{km}\gi^{pq}~\Delta~\tr_{qp}\tr_{mk}\Big){\bf \Psi}=0= \TR_{i(j}\TR_{kl)}{\bf \Psi} ~.
\eea
To construct irreps of $\mathfrak{gl}(d)$ starting from the representation in~(\ref{yt1}),
we introduce the operator ${\bf N}^2_{1}$ which moves a box from the
$1^{\!st}$ row to the $2^{nd}$ row, and note that the Lie algebra
$$
\mathfrak{g}=\Big\{
\GRAD^i,\Delta,{\bf N}^2_{1},\tr_{ij},\dv_i\Big\}\, ,
$$
is  first class since
$$
[{\bf N}_1^2,\GRAD^1]=\GRAD^2 ~,~~~
\Big[\begin{pmatrix}\TR_{12}\\\TR_{22}\end{pmatrix},{\bf N}_1^2\Big]=
\begin{pmatrix}\TR_{11}\\  2\TR_{12}\end{pmatrix}~,~~~
[\dv_2,{\bf N}_1^2]=\dv_1~.
$$
In particular, because ${\bf N}_1^2$ commutes with $\mathcal{G}$, gauging this operator produces a Dirac constraint ${\bf N}_1^2{\bf \Psi}=0$ which selects from  the tensor product (\ref{yt1}) an  irreducible $\mathfrak{gl}(d)$ representation:
\be 
\Yvcentermath1
{\bf N}_1^2{\bf \Psi}=0~~\rightarrow~~
{\bf \Psi}=\yng(4,3)~~~~~.
\ee



\subsection*{$\mathfrak{osp}(1,1|2)$: Mixed Multi-Forms}
Our final example is the theory of mixed multi-forms 
\be
\Yvcentermath1
{\bf \Psi}=~{\bf \Psi}_{\mu_1...\mu_{r_1}\nu_1...\nu_{s_1}}dx^{\mu_1}\wedge\ldots\wedge dx^{\mu_{r_1}}\!
\otimes
dx^{\nu_1}\odot\ldots\odot dx^{\nu_{s_1}}=~ \yng(1,1,1)~ \otimes  ~\yng(4) ~.
\ee
Explicitly we find, in an obvious notation, the following equation of motion
\bea
\mathcal{G}{\bf \Psi}&=&\Big(\Delta-\gi^{{\rm f}{\rm b}}\Delta\tr_{{\rm b}{\rm f}}-\frac 1 2 \gi^{{\rm b}{\rm b}}\Delta\tr_{{\rm b}{\rm b}}+\frac1 4 \gi^{{\rm f}{\rm b}}\gi^{{\rm f}{\rm b}}\Delta\tr_{{\rm b}{\rm f}}\tr_{{\rm b}{\rm f}} \nonumber\\
&&
 -\,{\bf d}\, (1-\frac 1 4\gi^{{\rm b}{\rm b}}\tr_{{\rm b}{\rm b}}-\frac{1}{12}\gi^{{\rm f}{\rm b}}\gi^{{\rm f}{\rm b}}\tr_{{\rm b}{\rm f}}\tr_{{\rm b}{\rm f}})~{\bf d}^* \nonumber\\
 &&
 -{\bf grad}~(1+\frac 1 4\gi^{{\rm b}{\rm b}}\tr_{{\rm b}{\rm b}}
 -\frac{1}{12}\gi^{{\rm f}{\rm b}}\gi^{{\rm f}{\rm b}}\tr_{{\rm b}{\rm f}}\tr_{{\rm b}{\rm f}})~{\bf div}\nonumber\\
 &&
 -\, \frac 1 2 {\bf d}\, \gi^{{\rm b}{\rm b}}\tr_{{\rm b}{\rm f}}~{\bf div}-\frac 1 2 {\bf grad}~\gi^{{\rm f}{\rm b}}\tr_{{\rm b}{\rm b}}~{\bf d}^*\nonumber\\
&&
+\,\gi^{{\rm f}{\rm b}}(1-\frac 1 2\gi^{{\rm f}{\rm b}}\tr_{{\rm b}{\rm f}})\dv~ {\bf d}^*
+{\bf d} \, \gr(1-\frac 1 2\gi^{{\rm f}{\rm b}}\tr_{{\rm b}{\rm f}})\tr_{{\rm b}{\rm f}}
\nonumber\\
&&+\,\frac 1 2 \gi^{{\rm b}{\rm b}}\dv^2+\frac 1 2\gr^2\, \tr_{{\rm b}{\rm b}}\Big){\bf\Psi}=0
=\TR_{\rm bb}^2{\bf\Psi}
=\TR_{\rm bb}\TR_{\rm bf}{\bf\Psi}\nn \, . 
\eea
Also in this case the irreps of $\mathfrak{gl}(d)$ can be  obtained gauging the operator~${\bf N}_{\rm f}^{\rm b}$ which selects Young tableaux of the shape
\be \nn 
\Yvcentermath1
{\bf N}_{\rm f}^{\rm b}{\bf \Psi}=0~~\rightarrow~~
{\bf \Psi}=\yng(4,1,1,1)~~~~~.
\ee

\section*{Conclusions and Outlook}

In this Letter we presented gauge invariant action principles 
for mixed multi-forms in terms of
  generating functions for  generalized Einstein operators. This gives a simple formulation of mixed symmetry higher spins along with a calculus for handling the operator expressions involved. 
The results were achieved using the  BRST detour quantization techniques of~\cite{Cherney:2009mf},
applied to the  $\mathfrak{osp}(q,q|2r)$ quantum mechanical models of~\cite{Hallowell:2005np}. Moreover we discussed two interesting examples in which we showed how to construct $\mathfrak{gl}(d)$ irreducible representations. These ideas are easily extended to general cases: irreducible $\mathfrak{gl}(d)$ representations are obtained by studying all possible permutation symmetries, to this end one can introduce  operators {${\bf N}^{j > i}_{i}$
which move a box from row $j$ to row $i$. Since these operators commute with our Einstein operators,
their vanishing may be imposed as an additional  set of Dirac constraints ${\bf N}^{j > i}_{i}{\bf \Psi}=0$ which select particular Young tableaux shapes
(a more detailed analysis of this construction my be found in~\cite{Skvortsov:2008vs,Alkalaev:2008gi}).
 
Throughout this Letter we worked on flat backgrounds, but the constraint algebra studied can also be represented by tensor operators on manifolds
with more interesting geometric structures. Our results can therefore be also applied to those cases. 
In particular, the $\mathfrak{so}(2,2)$ case  have been analyzed on K\"ahler manifolds in~\cite{Cherney:2009vg}  (K\"ahler spinning particle models were studied in~\cite{Marcus,Fiorenzo}). The extension of the Einstein generating functions to curved space is also an interesting task; in conformally flat backgrounds the $\mathfrak{osp}(q,q|2r)$  (super)algebras have quadratic corrections built from the Casimirs of their $R$-symmetry generators so, after minimal covariantization, one could try to construct curved space correction by requiring gauge invariance and adding non-minimal couplings\footnote{A representation theoretic analysis of this problem has been given in~\cite{Metsaev:2004ee}.}. A BRST detour quantization of that higher order algebra may also yield results for those cases (for BRST studies of (A)dS higher spins see~\cite{Buchbinder:2005cf}). 

Moreover, motivated by the observation that four-dimensional, supersymmetric,  black hole solutions to 
${\cal N}=2$ supergravities 
lead to spinning particles with 
${\cal N}=4$ worldline supersymmetry~\cite{Pioline:2008zz}, we can also  analyze those systems. In this case the BRST 
quantization involves first class constraint algebras with structure functions 
related to the underlying quaternionic K\"ahler structures of the spinning particle target spaces; we have found a 
geometric construction of the BRST charges of these models and will present 
their quantization, as well as a gauge invariant action principle, in a future work~\cite{future}.


\begin{thebibliography}{99}

\bibitem{Bekaert:2005vh}
  M.~A.~Vasiliev,
  Fortsch.\ Phys.\  {\bf 52}, 702 (2004)
  [arXiv:hep-th/0401177];
  D.~Sorokin,
  AIP Conf.\ Proc.\  {\bf 767}, 172 (2005)
  [arXiv:hep-th/0405069];
  N.~Bouatta, G.~Compere and A.~Sagnotti,
 [ arXiv:hep-th/0409068];
  X.~Bekaert, S.~Cnockaert, C.~Iazeolla and M.~A.~Vasiliev,
  [arXiv:hep-th/0503128].

\bibitem{Hallowell:2005np}
  K.~Hallowell and A.~Waldron,
  Nucl.\ Phys.\  B {\bf 724} (2005) 453
  [arXiv:hep-th/0505255];
  SIGMA {\bf 3}, 089 (2007), 
  [arXiv:0707.3164 [math.DG]];
  Commun.\ Math.\ Phys.\  {\bf 278} (2008) 775
  [arXiv:hep-th/0702033].
%

\bibitem{Bastianelli:2007pv}
  F.~Bastianelli, O.~Corradini and E.~Latini,
  JHEP {\bf 0702} (2007) 072
  [arXiv:hep-th/0701055];
  JHEP {\bf 0811} (2008) 054
  [arXiv:0810.0188 [hep-th]].


 \bibitem{Marcus}
N.~Marcus and S.~Yankielowicz, 
Nucl. Phys. {\bf B} 432 (1994) 225 [arXiv:hep-th/9408116], 
N. Marcus, 
Nucl. Phys. {\bf B} 439, 583 (1995) 
[arXiv:hep-th/9409175]. 



\bibitem{Cherney:2009mf}
  D.~Cherney, E.~Latini and A.~Waldron,
  [arXiv:0906.4814 [hep-th]].

\bibitem{Alkalaev:2008gi}
  K.~B.~Alkalaev, M.~Grigoriev and I.~Y.~Tipunin,
  Nucl.\ Phys.\  B {\bf 823}, 509 (2009)
  [arXiv:0811.3999 [hep-th]].


\bibitem{Labastida:1986gy}
  J.~M.~F.~Labastida and T.~R.~Morris,
  Phys.\ Lett.\  B {\bf 180}, 101 (1986).
  J.~M.~F.~Labastida,
  Phys.\ Rev.\ Lett.\  {\bf 58}, 531 (1987);
  Nucl.\ Phys.\  B {\bf 322}, 185 (1989).

\bibitem{de Medeiros:2002ge}
  P.~de Medeiros and C.~Hull,
  Commun.\ Math.\ Phys.\  {\bf 235}, 255 (2003)
  [arXiv:hep-th/0208155].



\bibitem{Henneaux:2008ee}
  M.~Henneaux,
  Int.\ J.\ Geom.\ Meth.\ Mod.\ Phys.\  {\bf 5} (2008) 1255
  [arXiv:0808.1975 [hep-th]];
  M.~Dubois-Violette and M.~Henneaux,
  Commun.\ Math.\ Phys.\  {\bf 226} (2002) 393
  [arXiv:math/0110088];
  Lett.\ Math.\ Phys.\  {\bf 49} (1999) 245
  [arXiv:math/9907135].


\bibitem{Bekaert:2002dt}
  X.~Bekaert and N.~Boulanger,
  Commun.\ Math.\ Phys.\  {\bf 245} (2004) 27
  [arXiv:hep-th/0208058];
  Commun.\ Math.\ Phys.\  {\bf 271} (2007) 723
  [arXiv:hep-th/0606198].

\bibitem{Bastianelli:2009eh}
  F.~Bastianelli, O.~Corradini and A.~Waldron,
  ``Detours and Paths: BRST Complexes and Worldline Formalism,''
  [arXiv:0902.0530 [hep-th]].
  
  
  
\bibitem{Skvortsov:2008vs}
  E.~D.~Skvortsov,
  JHEP {\bf 0807}, 004 (2008)
  [arXiv:0801.2268 [hep-th]];
  Nucl.\ Phys.\  B {\bf 808}, 569 (2009)
  [arXiv:0807.0903 [hep-th]].
  
\bibitem{Zinoviev:2002ye}
  Yu.~M.~Zinoviev,
  [arXiv:hep-th/0211233];
  ``First order formalism for mixed symmetry tensor fields,''
  [arXiv:hep-th/0304067];
   ``First order formalism for massive mixed symmetry tensor fields in
  Minkowski and (A)dS spaces,''
  [arXiv:hep-th/0306292].

\bibitem{Campoleoni:2008jq}
  A.~Campoleoni, D.~Francia, J.~Mourad and A.~Sagnotti,
  Nucl.\ Phys.\  B {\bf 815}, 289 (2009)
  [arXiv:0810.4350 [hep-th]].


\bibitem{Campoleoni:2009gs}
  A.~Campoleoni, D.~Francia, J.~Mourad and A.~Sagnotti,
  arXiv:0904.4447 [hep-th].


\bibitem{Burdik:2000kj}
  C.~Burdik, A.~Pashnev and M.~Tsulaia,
  Nucl.\ Phys.\ Proc.\ Suppl.\  {\bf 102}, 285 (2001)
  [arXiv:hep-th/0103143]; Mod.\ Phys.\ Lett.\  A {\bf 16}, 731 (2001)
  [arXiv:hep-th/0101201].



\bibitem{Cherney:2009vg}
  D.~Cherney, E.~Latini and A.~Waldron,
  Phys.\ Lett.\  B {\bf 674} (2009) 316
  [arXiv:0901.3788 [hep-th]].

\bibitem{Henneaux:1987cp}
  M.~Henneaux and C.~Teitelboim,
  ``First And Second Quantized Point Particles Of Any Spin,''
 In Santiago 1987, Proceedings, ``Quantum mechanics of fundamental systems'' {\bf 2} 113.

 

\bibitem{Fiorenzo}
F.~Bastianelli and R.~Bonezzi, ``$U(N)$ spinning particles and higher spin equations on complex manifolds'',
[arXiv:0901.2311[hep-th]].

  
\bibitem{Metsaev:2004ee}
  R.~R.~Metsaev,
  Class.\ Quant.\ Grav.\  {\bf 22}, 2777 (2005)
  [arXiv:hep-th/0412311].


\bibitem{Buchbinder:2005cf}
  I.~L.~Buchbinder and V.~A.~Krykhtin,
  [arXiv:hep-th/0511276].
  I.~L.~Buchbinder, V.~A.~Krykhtin and H.~Takata,
  Phys.\ Lett.\  B {\bf 656} (2007) 253
  [arXiv:0707.2181 [hep-th]].
  I.~L.~Buchbinder, V.~A.~Krykhtin and A.~Pashnev,
  Nucl.\ Phys.\  B {\bf 711} (2005) 367
  [arXiv:hep-th/0410215].
  I.~L.~Buchbinder, A.~Pashnev and M.~Tsulaia,
  Phys.\ Lett.\  B {\bf 523} (2001) 338
  [arXiv:hep-th/0109067].

\bibitem{Pioline:2008zz}
  B.~Pioline,
  Lect.\ Notes Phys.\  {\bf 755}, 1 (2008);
  Class.\ Quant.\ Grav.\  {\bf 23}, S981 (2006)
  [arXiv:hep-th/0607227].
  














  
  
  
%








  
  
 






\bibitem{future}
  D.~Cherney, E.~Latini and A.~Waldron, in preparation.


\end{thebibliography}
\end{document}